# Blockchain-assisted Undisclosed IIoT Vulnerabilities Trusted Sharing Protection with Dynamic Token

Wenbo Zhang[1], Jing Zhang[1], Yifei Shi[1], and Jingyu Feng[1]

[1] School of Cyberspace Security, Xi'an University of Posts & Telecommunications, Xi'an 710121, China.
*Corresponding author: Jingyu Feng, E-mail: fengjy@xupt.edu.com.

*Abstract*—With the large-scale deployment of industrial internet of things (IIoT) devices, the number of vulnerabilities that threaten IIoT security is also growing dramatically, including a mass of undisclosed IIoT vulnerabilities that lack mitigation measures. Coordination Vulnerabilities Disclosure (CVD) is one of the most popular vulnerabilities sharing solutions, in which some security workers (SWs) can develop undisclosed vulnerabilities patches together. However, CVD assumes that sharing participants (SWs) are all honest, and thus offering chances for dishonest SWs to leak undisclosed IIoT vulnerabilities. To combat such threats, we propose an Undisclosed IIoT Vulnerabilities Trusted Sharing Protection (UIV-TSP) scheme with dynamic token. In this article, a dynamic token is an implicit access credential for an SW to acquire an undisclosed vulnerability information, which is only held by the system and constantly updated as the SW access. Meanwhile, the latest updated token can be stealthily sneaked into the acquired information as the traceability token. Once the undisclosed vulnerability information leaves the SW host, the embedded self-destruct program will be automatically triggered to prevent leaks since the destination MAC address in the traceability token has changed. To quickly distinguish dishonest SWs, trust mechanism is adopted to evaluate the trust value of SWs. Moreover, we design a blockchain-assisted continuous logs storage method to achieve the tamper-proofing of dynamic token and the transparency of undisclosed IIoT vulnerabilities sharing. The simulation results indicate that our proposed scheme is resilient to suppress dishonest SWs and protect the IoT undisclosed vulnerabilities effectively.

## I. Introduction

The Internet of Things (IoT) technology is being applied to every part of our lives [1]. As a subset of IoT, Industrial Internet of Things (IIoT) has recently attracted attention [2]-[4]. By leveraging sensors, actuators, GPS devices and mobile devices. the Internet of Things technology is being applied to advance the development of many industrial systems [5]. The industrial systems where this IIoT technology is integrated include energy [6], manufacturing[7], logistics[8] and transportation [9], and so on.

Currently, IIoT devices have been widely deployed with weak security features or a lack of security [10]. These features have made IIoT devices a good target for attackers with malicious intentions, and in many cases, exploits using IIoT devices have been occurring[11]. There is an urgent need for a solution that provides a lightweight and low-cost mechanism for collaborative security response of IIoT devices against emerging vulnerabilities [12].

However, the IIoT vendors generally have weak security emergency response capabilities. It is better to invite some security workers (such as organizations, institutions, or white hats) to help them path the new vulnerabilities of IIoT devices. In order to standardize the process of vulnerabilities patching and accelerate the development of mitigation measures, the vulnerabilities disclosure policy has been presented in [13], including vulnerabilities reporting, sharing, coordinating, and patching. An IIoT vulnerability can be officially disclosed after the patch is made, otherwise it is called an undisclosed IIoT vulnerability (*uiv*). According to the vulnerabilities disclosure policy, the IIoT vendors can report a new *uiv* and share it with some security workers (SWs) who develop their patches together by means of the Coordination Vulnerabilities Disclosure (CVD).

Unfortunately, CVD is a set of guidelines without mandatory measures [14][13]. CVD assumes that SWs are all honest, and thus offering chances for dishonest SWs to leak undisclosed IIoT vulnerabilities. Due to the widespread use of IIoT devices, the leakage of an *uiv* information could cause a large-scale damage. Therefore, it is necessary to prevent the leakage of the *uiv*. Although a lot of works have been done in the field of threat intelligence sharing [16]-[20] they focus on the sharing protection of disclosed vulnerabilities information. Little attention has been paid to the sharing protection of *uiv* information.

In this article, we propose an Undisclosed IIoT Vulnerabilities Sharing Protection (UIV-TSP) scheme with dynamic token. Our final objective is to prevent undisclosed IIoT vulnerabilities leakage until their patches are released. The main contributions of this article are as follows:

1) Introduce dynamic token as the implicit access credential and traceability clue for an SW. When a SW uploads a new *uiv*, each internal SW is assigned a corresponding token called $token_{access}$, which is only held by the system and cannot be seen by anyone. Even if an SW is granted access through identify authentication, the *uiv* information cannot be acquired without $token_{access}$. To avoid malicious inference, $token_{access}$ should be updated dynamically. At the end of SW access, the current $token_{access}$ is revoked. A new random number is integrated into the hash generation of token to get a new $token_{access}$ as the next access credential. Meanwhile, the current $token_{access}$ and the MAC address of the SWs are hashed to create the traceability token called $token_{tracing}$, which is embedded in the undisclosed IIoT vulnerability acquired by the SW.

2) Design a blockchain-assisted method to store the

continuous logs of all SWs for the UIV-TSP scheme. To ensure the tamper-proofing of dynamic token, the original token and its all subsequent updates should be stored on the blockchain. To achieve the transparency of *uiv* sharing, their metadata and the related SW access records are also stored on the blockchain.

3) Present a leakage prevention method with one-step traceability. A benign logic bomb called $code_{preleak}$ is embedded into the *uiv* information, which checks that whether the current MAC address is the same as the preset destination address in $token_{tracing}$. Due to confidentiality, the *uiv* information can only be reached by one-step to the SW host that is licensed by the system. Once the *uiv* information leaves the SW host, $code_{preleak}$ will automatically trigger the self-destruct program to prevent leaks.

4) Adopt trust mechanism to evaluate the trust value of SWs according to their historical behaviors. With high trust value, honest SWs would be accepted to acquire undisclosed IoT vulnerabilities information. With low trust value, dishonest SWs would be rejected. With medium trust value, it is difficult to determine the access authorities of semi-honest SWs. In this case, we can release a false *uiv* with $token_{tracing}$ to trap their conspirators.

The architecture of this paper is as follows. In Section II, we introduce the related works. Our UIV-TSP scheme is proposed in Section III. We perform the simulation analysis of our UIV-TSP scheme in Section IV. We also discuss the industrial application prospect of our UIV-TSP scheme in Section V. Finally, we conclude the article in Section VI.

## II. RELATED WORKS

ISO/IEC 29147 defines vulnerability disclosure is a process through which vendors and vulnerability finders may work cooperatively in finding solutions that reduce the risks associated with a vulnerability [13]. In security practices, the vulnerability disclosure often falls into two extremes [14], including 1) Public Disclosure: a vulnerability will be fully disclosed as soon as reported, 2) Private Disclosure: the vulnerability will remain confidential and undisclosed to the public. But, the first case cannot protect against unpatched vulnerabilities. In the second case, if the vulnerability is detected by someone else, it can give a fatal blow to IIoT devices if necessary. Currently, Coordination Vulnerabilities Disclosure (CVD) [14] is a good choice to develop undisclosed vulnerabilities patches together among some security workers. As shown in Fig. 1, the CVD process is consisted of gathering information from a vulnerability finder, coordinating the sharing information among relevant stakeholders, and disclosing the existence of vulnerabilities and their mitigations to various stakeholders [14].

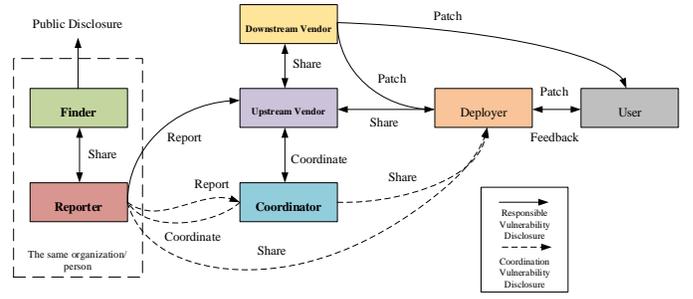

Fig. 1.Flowchart of vulnerabilities disclosure policy[14]

To ensure quick response when the same incident occurs, the threat intelligence sharing containing *uiv* has been presented. Threat intelligence is defined as the process of obtaining many sources and knowledge about cyber threats that can be used for discovering malicious events for the purpose of protecting organizations' assets [15]. To prevent the leak of sensitive data in threat intelligence, many studies have integrated cryptography primitives into their threat intelligence sharing scheme. Vakilinia *et al.*[16] design a mechanism enables the organizations to share their cybersecurity information anonymously. Meanwhile, they propose a new blind signature based on BBS+ to reward contributions anonymously. Badsha *et al.* [17] propose a privacy preserving protocol where organizations can share their private information as an encrypted form with others and they can learn the information for future prediction without disclosing any private information. Fuentes *et al.* [18] introduce PRACIS, a scheme for cybersecurity information sharing that guarantees private data forwarding and aggregation by combining STIX and homomorphic encryption primitives. Homan *et al*.[19] leverage the security properties of blockchain and design a more effective and efficient framework for cybersecurity information sharing network. Davy *et al*.[20] employ blockchain and CP-ABE to offer fine-grained protection and trustworthy threat intelligence sharing with the ability to audit the provenance of threat intelligence.

However, these schemes are helpful to protect the *uiv* information sharing, while the sharing protection of undisclosed vulnerabilities information has not been involved. Without mitigation measures, the leakage of an undisclosed IIoT vulnerability could cause a large-scale damage. Therefore, while protecting the sharing of undisclosed vulnerabilities information, the responsibility of SWs should be traced back.

Currently, some responsibility traceability schemes have been presented. Leo *et al*. [21] propose a fair traitor tracing scheme to secure media sharing in the encrypted cloud media center by proxy re-encryption and fair watermarking. Ning *et al*. [22] present a traitor tracing with CP-ABE scheme by two kinds of non-interactive commitments. In terms of tracing data leakage, some researchers focus on leveraging cryptography algorithms to implement an accountable and efficient data sharing. Mangipudi *et al*. [23] present an accountable data sharing scheme to fairly track the traitor of leaked data by Oblivious Transfer (OT) protocol and zero knowledge (ZkPok).Cheng *et al*. [24] design an accountable and efficient data sharing scheme for industrial IoT (IIoT), named ADS, in which data receiver's private key (i.e., evidence) is embedded in sharing data, data owner can pursue the responsibility of a public for profits while without permission. Nevertheless, these schemes cannot deal well with the trade-off between traceability time and robustness. A lightweight traceability scheme is required to *uiv*.

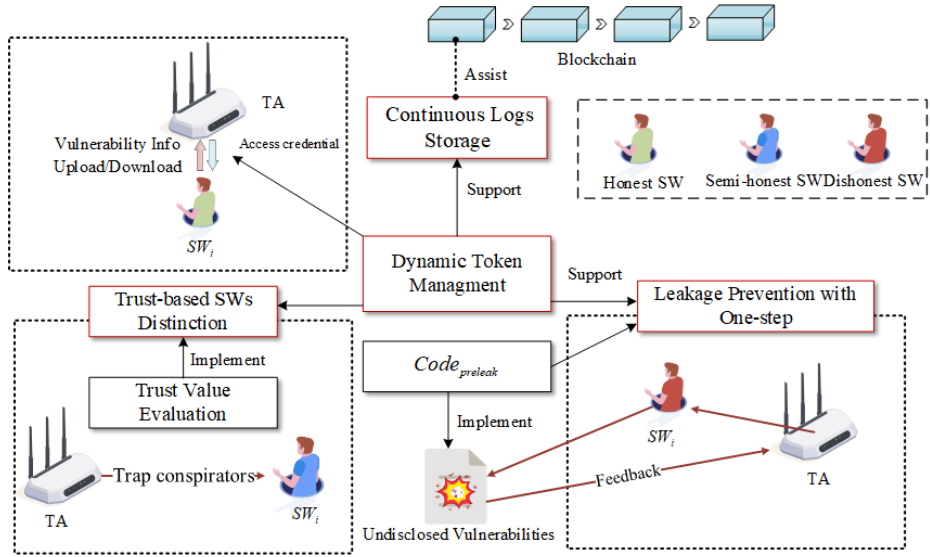

Fig. 2. System architecture of UIV-SP scheme

## III. OUR PROPOSED UIV-TSP SCHEME

With dynamic token, we propose an Undisclosed IIoT Vulnerabilities Sharing Protection scheme called UIV-TSP to prevent *uiv* leakage until their patches are released. Concretely, the UIV-TSP scheme consists of four collaborative modules: dynamic token management, blockchain-assisted continuous logs storage, leakage prevention with one-step traceability, and trust-based SWs distinction.

### A. System Architecture

To prevent *uiv* leakage, we first present the system architecture of the UIV-TSP scheme. Fig. 2 illustrates the overall system architecture of our scheme, which consists of the following entities:

1) Trusted Authority (TA): Trusted authority is responsible for processing SW's access requests, generating and updating dynamic token, and evaluate trust value of $SW_i$.

2) Security Workers (SWs): There exist some workers who access *uiv* information in the sharing environment to develop their mitigation measures. We describe three types of workers: 1) Honest SWs who does not engage in unauthorized access;2) Semi-honest SWs who have a chance of committing malicious behavior; 3) Dishonest SWs who often leak *uiv* information. In this article, SWs are defined as $\mathcal{W} = \{SW_1, SW_2 \ldots SW_m\}, SW_i \in \mathcal{W}$.

### B. Dynamic Token Management

We introduce dynamic token as the implicit access credential and traceability clue for an SW. As show in Fig. 3, the lifecycle of dynamic token is consisted of generation and update.

*1) Token Generation*

When a SW submits a new undisclosed vulnerability $vul_j$, each internal SW is assigned a corresponding token called $token_{access}$, which is only held by the system and cannot be seen by anyone. TA generates the initial $token_{access}$ through a hash function. Meanwhile, we define $vul_{meta}$ is the meta information of an *uiv*. The initial $token_{access}$ can be calculated as:

$$token_{access} = H(SW_i||vul_{meta}||tp||nonce) \quad (1)$$

Even if an SW is granted access through identify authentication, the *uiv* cannot be acquired without $token_{access}$. Algorithm 1 is performed to generate $token_{access}$.

---
**Algorithm 1** Token generation
---
**Input**: $SW_i$
**Output**: $token_{access}$
1: **If** $SW_i$ in SWs pool **then**
2:   $token_{access} = H(SW_i||vul_{meta}||tp||nonce)$
3:   **If** $token_{access}$ in blockchain **then**
4:     Update $token_{access}, SW_i \leftrightarrow token_{access}$
5:   **else**
6:     Store $token_{access}, SW_i \leftrightarrow token_{access}$
7:   **end if**
8: **end if**
---

*2) Token Update*

To avoid malicious inference, we integrate a one-time random number into the hash generation of token to update token dynamically. At the end of SW access, $token_{access}$ can be update as:

$$token_{access} \leftarrow H(SW_i||vul_{meta}||tp||nonce') \quad (2)$$

Meanwhile, the current $token_{access}$ and the MAC address of the SW are hashed to create the traceability token called $token_{tracing}$. $token_{tracing}$ can be defined as:

$$token_{tracing} \leftarrow H(token_{access}||mac_{current}) \quad (3)$$

In our scheme, $token_{tracing}$ can be stealthily sneaked into the acquired $vul_j$ information as the traceability credential.

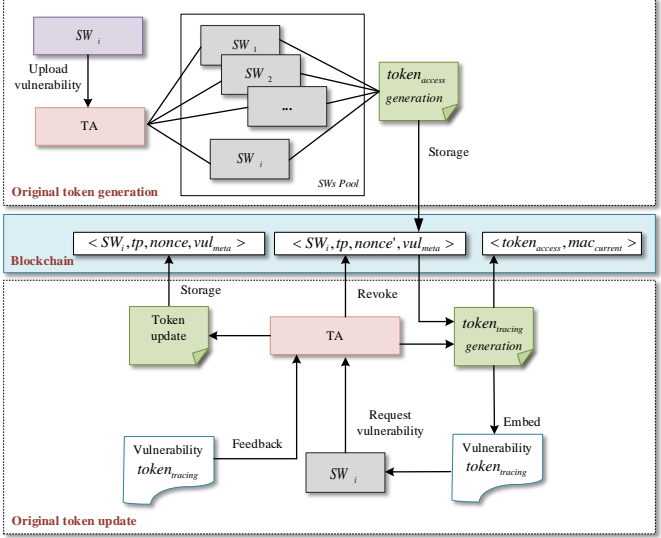

Fig. 3. Lifecycle of dynamic token

Algorithm 2 is performed to update $token$ and embed it into the acquired $vul_j$ information.

---
**Algorithm 2** Token update and embed
**Input:** $R[i]$
**Output:** $token_{access}$ and $token_{tracing}$
1: **If** end of request access **then**
2: revoke $token_{access}$
3: $token_{tracing} \leftarrow H(token_{access}||mac_{current})$
4: New $token_{access} = H(SW_i||vul_{meta}||tp||nonce')$
5: Store $token_{access}, SW_i \leftrightarrow token_{tracing}$
6: Embed $token_{tracing}$ to $vul_j$
7: **end if**

---

As shown in Fig. 4, the execution strategies of dynamic token management can be executed with four steps.

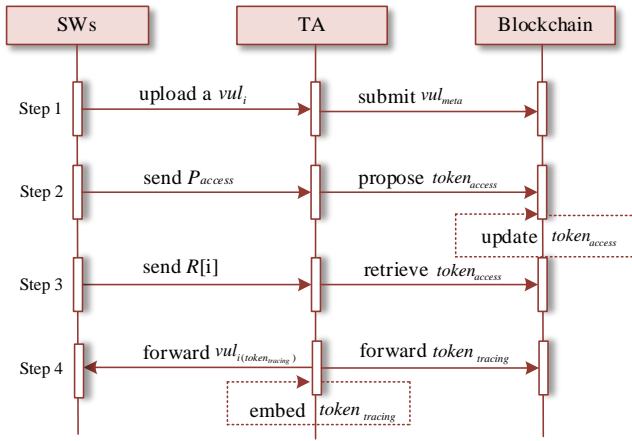

Fig. 4. Execution strategies of dynamic token management

Step1. After receiving the undisclosed vulnerability $vul_j$ from SW, TA will submit the meta information of an undisclosed vulnerability $vul_{meta}$ to blockchain.

Step2. SW sends the access list ($P_{access}$) about the $vul_j$ to determine which an SW can access. TA generates $token_{access}$ and sends it to the blockchain. TA will update $token_{access}$ and revoke the old $token_{access}$ before the next access.

Step3. An SW sends a vulnerability access request $R[i]$. TA will retrieve his $token_{access}$ in the blockchain.

Step4. TA responds the access request $R[i]$, and then generates $token_{tracing}$ which is embedded into the acquired $uiv$. The $token_{tracing}$ is also stored to the blockchain. The SW will get the $uiv$ embedded in the traceability token.

### C. Blockchain-assisted Continuous Logs Storage

Blockchain was first introduced as the technology supporting Bitcoin [25]. Due to the advantages including transparency, traceability, and tamper-proofing, many studies have integrated blockchain into their scheme to implement a reliable and efficient data storage [26], [27].

In general, there are three types of blockchain data storage patterns [28]:

1) Public blockchain: A public blockchain is the blockchain that can read by anyone in the world, anyone can send transactions to and expect to see them included, if they are valid, and anyone can participate in the consensus process, which determines what blocks get added to the chain and what the current state is.

2) Private blockchain: A private blockchain is the blockchain where can write by only one organization.

3) Consortium blockchain: A consortium blockchain is the blockchain where the consensus process is controlled by a pre-selected set of miners.

Since token can only be held by the system, the public blockchain is obviously cannot meet the requirements. Hence, a private blockchain-assisted storage method is designed to centralized storage logs. To prevent attackers from tampering with data, the logs of SWs' activities in the shared process should also be continuously recorded on the blockchain. These continuous logs, including dynamic token, trust value of SW ($Tr_i$), and behaviors record ($R[i]$), are also only held by the system. It can be found that the private blockchain is suitable for our scheme.

In the blockchain, the block structure of continuous logs storage is shown in Fig. 5.

#### 1) Block Head

The block head is slightly different from the traditional structure. Except for previous hash, timestamp, Merkle Root and block ID, several new elements are integrated into the block head:

- $SW_i$: The ID of the $i$-th SW who requests the undisclosed vulnerability in the sharing environment.
- $Tr_i$: The trust value of $SW_i$. In the block head, $Tr_i$ can be quickly retrieved by TA.
- $vul_{meta}$: The meta information of an undisclosed IoT vulnerability.

#### 2) Block Body

In the block body, the logs data of an SW (such as $SW_i$) are hashed to build the Merkle tree. Except for $token_{access}$ and $token_{tracing}$, the log data of $SW_i$ contains the following

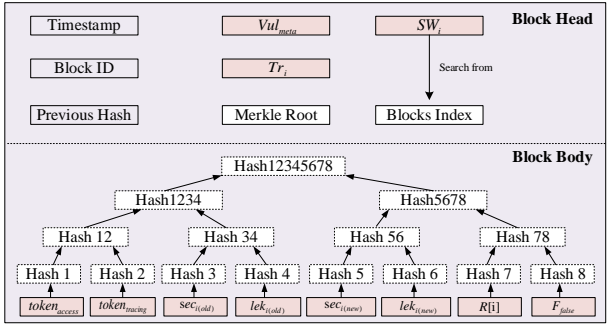

Fig. 5. Block structure of continuous logs storage

elements:
- $(sec_{i(old)}, lek_{i(old)})$: The historical trust data of $SW_i$, which can be used to evaluate the trust value of $SW_i$ before the access.
- $(sec_{i(new)}, lek_{i(new)})$: The current trust data of $SW_i$, which can be used to update the trust value of $SW_i$ after the access.
- $R[i]$: The access request record of $SW_i$ to an undisclosed vulnerability information.
- $F_{false}$: The flag whether a false *uiv* information has been released.

*D. Leakage Prevention Method with One-step Traceability*

In order to prevent SWs leaking the acquired $vul_j$ information, $vul_j$ should be self-destruct when they leave the host of SWs one-step. Thus, we design a benign self-triggering logic bomb $code_{Preleak}$.

$code_{Preleak}$ is composed of trigger condition and response payloads. The trigger condition is designed to detect the access environment difference between honest SWs and dishonest SWs, so that the protection payload will be activated to destroy $vul_j$ on the leakage side.

As shown in Fig. 6, the functional structure of $code_{Preleak}$ is consisted of self-checking and self-destruct.

*1) Self-Check*

Once the *uiv* information enters the SW host, $code_{preleak}$ will extract the current SW host Mac address and the revoked token to compute the verification value. Then, $code_{preleak}$ will match the verification value to the $token_{tracing}$. If the result $V_c$ is inconsistent, it will trigger self-destruct. $V_c$ can be calculated as:

$$V_c \leftarrow token_{tracing} == H(token_{access}, mac_{current})? \quad (4)$$

*2) Self-Destruct*

If $V_c=0$, the leakage unhappens. That is, the *uiv* information has not left the SW host. The protection payload continues to lurk.

If $V_c=1$, it may leak $vul_{id}$. That is, the *uiv* information has left the SW host. In this case, the protection payload will be activated immediately to destroy $vul_j$.

Algorithm 3 is performed to match verification value.

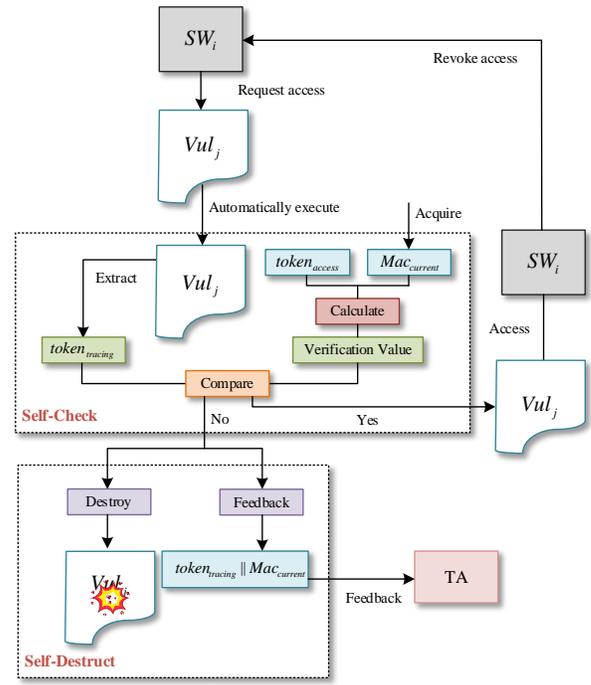

Fig. 6. Functional structure of $code_{preleak}$

---
**Algorithm 3** Match verification value

**Input**: $vul_j$
**Output:** $access$ or deny
1: $token_{tracing} = Extract(vul_j)$
2: **If** $token_{tracing}$ no existence **then**
3:    Go to step10
4: **else**
5:    Acquire $mac_{current}$
6:    $V_c \leftarrow token_{tracing} == H(token_{access}, mac_{current})?$
7:    **If** $V_c = 0$ **then**
8:      Assign access authorities
9:    **else**
10:     Invoke self-destruct
11:   **end if**
12: **end if**

---

Algorithm 4 is performed to activate the protection payload.

---
**Algorithm 4** Activate protection payload

**Input**: $vul_j$
**Output:** destroy $vul_j$
1: **If** $Extract(vul_j) == F_{false}$ **then**
2:   Go to step 6
3: **else**
4:   Send an encrypted feedback message
$e_f = \{token_{tracing}, vul_j, SW_i, mac_{current}, t_{feedback}\}$
to TA
5:   Destroy the target $vul_j$
6: **end if**

---

At the same time as the self-destruct event, $code_{Preleak}$ can automatically send an encrypted feedback $e_f = \{token_{tracing},$

$vul_j, SW_i, mac_{current}, t_{feedback}\}$ to TA. On the receipt of $e_f$:
- $vul_j$: The ID of undisclosed IoT vulnerability.
- $t_{feedback}$: The time to $code_{preleak}$ send the feedback message.

### E. Trust-based SWs Distinction

Trust mechanism can be adopted to evaluate the trust value of SWs according to their historical behaviors. With trust value, we can quickly distinguish honest SWs and dishonest SWs. For semi-honest SWs, we can further validate their credibility by tracing whether they have conspirators. Different SWs will gain different access authorities to acquire *uiv* information.

*1) Trust Value Evaluation*

In the process of vulnerabilities information sharing, the behaviors of an SW are generally dualistic: secret-keeping and leakage. With trust mechanism, if an SW often leaks information, he will get a low trust value.

To quantify these duality behaviors, the beta function is one of the most popular trust value evaluation methods. It first counts the number of secret-keeping and leakage by an SW, and then calculates the trust value with beta function denoted by $Beta(\alpha, \beta)$[29].

$$Beta(\alpha, \beta) = \frac{\Gamma(\alpha, \beta)}{\Gamma(\alpha)\Gamma(\beta)} \theta^{\alpha-1}(1-\theta)^{\beta-1} \quad (5)$$

where $\theta$ is the probability of duality behaviors, $0 \leq \theta \leq 1$, $\alpha > 0, \beta > 0$.

Take $SP_i$ as an example, $sec_i$ and $lek_i$ denote the number of secret-keeping and leakage in the sharing environment, respectively. The base trust value of $SW_i$ can be evaluated with beta function as:

$$BT_i = Beta(sec_i + 1, lek_i + 1) \quad (6)$$

Moreover, consider the condition $\Gamma(n) = (n-1)!$ When $n$ is an integer [30]. The expectation value of the beta function can be calculated as: $E[Beta(\alpha, \beta)] = \alpha/(\alpha + \beta)$. Considering that the trust value is limited in the interval [0,1], $BT_i$ can be further described as follows:

$$BT_i = \frac{1 + sec_i}{2 + sec_i + lek_i} \quad (7)$$

When $sec_i \geq 1$ and $lek_i = 0$, $BT_i$ is always calculated as 1. The SW is completely trusted under this condition.

Obviously, the base trust value that decays too slowly will give dishonest SWs more opportunities to leak. So, it is very essential to introduce a penalty factor, which can be calculated as:

$$P_i = e^{-\frac{sec_i + lek_i}{sec_i}} \quad (8)$$

With the punishment of $P_i$ to $BT_i$, the trust value of SW can be further evaluated as:

$$Tr_i = BT_i \cdot e^{-\frac{sec_i + lek_i}{sec_i}} \quad (9)$$

*2) Distinction Rules*

SWs can be split into honest, dishonest and semi-honest on the basis of their trust value. ($\delta_h$, $\delta_l$, $\delta_m$) are respectively set as the threshold of high, low and medium trust value. The specific distinction rules are as follows:

**R1:** For $Tr_i > \delta_h$, the $SW_i$'s access request for an *uiv* information will be accepted. In this situation, $SW_i$ is classified as honest.

**R2:** For $Tr_i < \delta_l$, the $SW_i$'s access request for an *uiv* information will be rejected. In this situation, $SW_i$ is classified as dishonest.

**R3:** For $\delta_l < Tr_i < \delta_m$, it's difficult to determine the access authorities of $SW_i$. In this situation, $SW_i$ is classified as semi-honest.

To validate the credibility of semi-honest SWs, we can trace whether they have conspirators. Let $\mu_i(\cdot)$ denote number of conspirators of $SW_i$.

If $\mu_i = 0$, there are no conspirators. Hence, $SW_i$ is temporarily considered as honest.

If $\mu_i \geq 1$, $SW_i$ may have several conspirators. The trust value of $SW_i$ will be set to 0. As a result, $SW_i$ will be removed from the CVD and barred from rejoining.

*3) Trap Conspirators*

If $SP_i$ is a semi-honest SW, a false *uiv* information can be released to trap his conspirators. This false *uiv* information is set to a valid time.

Within the valid time, $code_{Preleak}$ will not trigger the protection payload and provide the feedback messages which can be employed to build a set of leak path $Path_i$. Once a conspirator is trapped, $SW_i$ can be regarded as dishonest.

After the valid time, the protection payload is activated to destroy the false *uiv* information, so as not to spread too widely.

Algorithm 5 is performed to trap conspirators.

---

**Algorithm 5** Trap conspirators

**Input:** $SW_i$
**Output:** $Path_i$
1: Init trust value of SW
2: Init $Path_i$
3: $Tr_i$ evaluate based on the feedback $e_f$
4: SW send $R[i]$, retrieves SW historical trust values $Tr_i$
5: **If** $\delta_l < Tr_i < \delta_m$ then
6:    Release false $vul_i$, $F_{false}$=true
7: **else**
8:    Release $vul$
9:    $SW_i \leftrightarrow token_{tracing}$
10:    $\mu_i$++
11:    **If** *valid time* expire **then**
12:      Invoke Self-Destruct
13:    **else**:
14:      **If** $mac_{current}$ not in $Path_i$ **then**
15:        $Path_i.add(mac_{current})$
16:      **end if**
17:    **end if**
18: **end if**
19: **end for**

## IV. SIMULATION ANALYSIS AND DISCUSSION

### A. Simulation Setup

We perform simulations to validate the performance of the UIV-TSP scheme in Python 3.7. The default simulation elements are shown in TABLE I.

TABLE I.
DESCRIPTION OF SIMULATION ELEMENTS

| Parameters | Description | Default |
|---|---|---|
| $SP_i$ | Number of SWs | 2000 |
| $Per$ | Percentage of dishonest SWs | 10%-50% |
| $\delta$ | Threshold of trust value | (0.2,0.5,0.8) |
| $cycle$ | Number of cycle simulation | 200 |
| $\epsilon$ | Embedded times of access credential | (1,2,3,4) |
| $k$ | The length of access credential | (256,512,1024) |

The simulations are executed in cycle-based manner. At each cycle, a certain percentage of SWs are randomly selected as dishonest SWs. The behavioral pattern of honest SW is modeled to always keep secret, while dishonest SWs may leak an uiv information sometimes. Without punishment, dishonest SWs will hinder the establishment of a trusted vulnerability information sharing environment. In this section, the performance and overload of the proposed UIV-TSP scheme are experimentally evaluated under various settings.

### B. Simulation Result

We first analyze the detection and false alarm rate of our UIV-TSP scheme. (i.e., probability of successful detection leaker). To analyze the effectiveness better, we compare UIV-TSP with the traditional Undisclosed IoT Vulnerabilities Sharing Protection (UIV-SP) scheme without trust mechanism.

In this simulation, the detection rate of UIV-TSP is better than UIV-SP in Fig. 7, while the false alarm rate of UIV-TSP is lower than UIV-SP in Fig. 8. Therefore, our designed trust mechanism can improve the performance of UIV-SP distinctly in the construction of trusted sharing environment.

Then, we validate the performance of our UIV-TSP scheme against dishonest SWs, in terms of suppressing leakage behaviors.

Dishonest SWs will leak uiv in the sharing environment. Consequently, some leakage behaviors may generate in each cycle which may cause unnecessary waste of network resources. So, the first performance indicator of UIV-TSP is to suppress these malicious behaviors. As shown in Fig. 9, it is obvious that UIV-TSP is better than UIV-SP in suppressing leakage behaviors. This means that the trust mechanism plays a key role in detection of dishonest SWs. In this simulation, $\delta$ is set as (0.2,0.5,0.8) respectively.

We also analyze the uiv leakage prevention performance of UIV-TSP in terms of leakage probability. In this simulation, the percentage of dishonest SWs is set as 30% respectively. As shown in Fig. 10. the uiv leakage probability of UIV-TSP is also lower than UIV-SP.

Finally, we evaluate the traceability performance of UIV-TSP in terms of computational complexity. We count the number of time-consuming operations such as the symmetric-key encryption/decryption (SKE), public-key encryption/decryption (PKE), the cryptographic hash function (HASH), the exponential operation (EXP) in $G_q$, the multiplicative operation (MUL) in $G_q$. As shown in Table

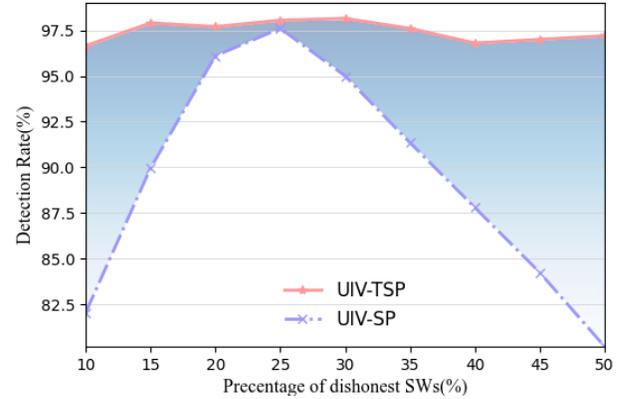

Fig. 7. Detection rate under the percentage of dishonest SWs

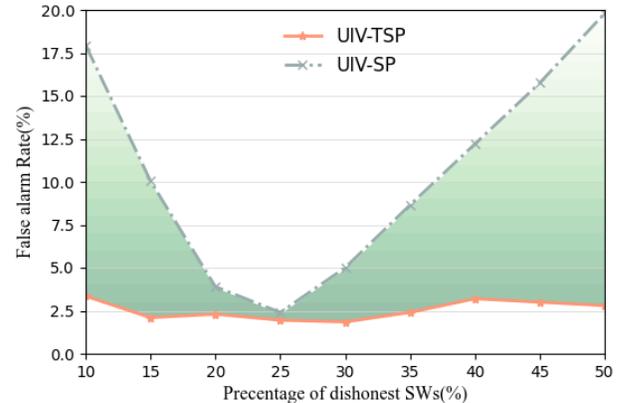

TABLE II., we compare UIV-TSP with three types of traditional schemes.

Fig. 8. Alarm rate under the percentage of dishonest SWs

It can be found that UIV-TSP cannot require any exponential operation or public-key encryption/decryption. Moreover, the requirements for hash and symmetric key operations are limited in UIV-TSP.

TABLE II.
COMPUTATIONAL COMPLEXITY

| | SKE | HASH | PKE | EXP | MUL |
|---|---|---|---|---|---|
| CROT[23] | 4t | 8t | 3t | 6t+3 | 2t+1 |
| ADS[24] | 2t | 2t | N/A | N/A | N/A |
| R-ADS[24] | 2t | 2t | N/A | N/A | 4kt+2t |
| Ours work | N/A | 3t | N/A | N/A | N/A |

To further evaluate the traceability performance of UIV-TSP in terms of computational complexity, we can observe the traceability delay of UIV-TSP and these traditional schemes. We run 200 rounds of experiments and obtain their average traceability delay as the result. We define $k$ is the length of access credential. In our UIV-TSP scheme, the access credential is a dynamic token. In the traditional schemes, the access credential is the private key of a user. A sufficient length of $k$ can contribute to the collision resistance generated by hashing. As the length of $k$ increases, the user capacity will be improved.

Of course, with the increase of $k$ and the embedded times of access credential, the traceability delay grows as well. As shown in Fig. 11, UIV-TSP is more computationally efficient than ADS and CROT. The reason is that the sharing data in UIV-TSP does not need to perform Oblivious Transfer (OT) and Zero-knowledge proof. Once the embedded times vary from 1 to 4, the number of OT increases linearly.

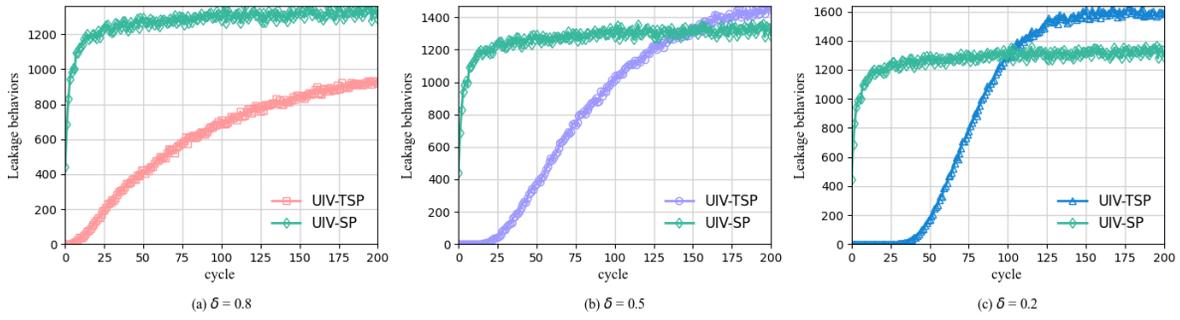

Fig. 9. Suppressing leakage behaviors

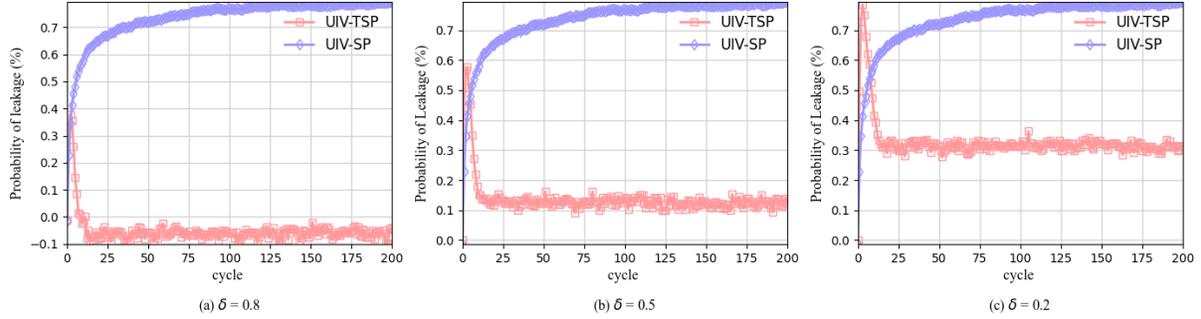

Fig. 10. Probability of leakage at different $\delta$

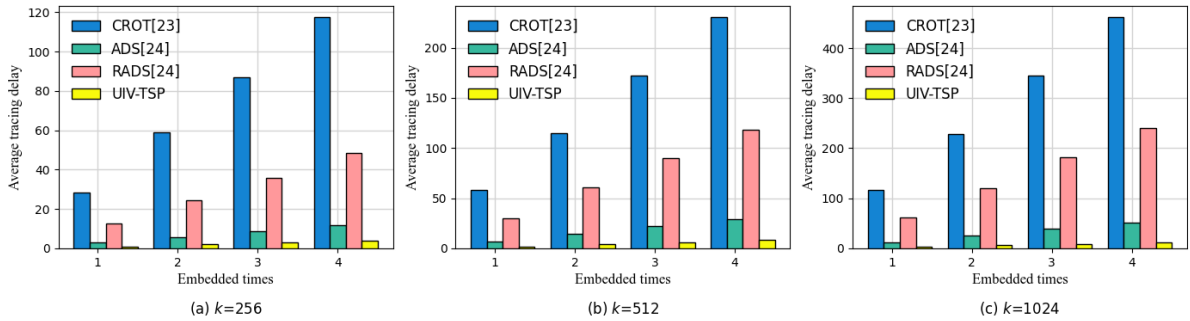

Fig. 11. Average tracing delay

In summary, our UIV-TSP scheme can prevent the *uiv* information leakage effectively, which merely requires limited traceability delay caused by multiple shared SWs.

## V. INDUSTRIAL APPLICATIONS DISCUSSION

Since the IIoT vendors generally have weak security emergency response capabilities, some security workers can be invited to help them path a new *uiv* by means of CVD. Our UIV-TSP scheme can prevent the *uiv* information leakage effectively in CVD and trace the dishonest SWs with limited traceability delay. As shown in Fig. 12, UIV-TSP can be applied to several IIoT scenarios, such as energy, logistics, manufacturing, and transportation, and so on.

Take the IIoT manufacturing as an example. Once an IIoT manufacturing vendor reports a new *uiv* in CVD, he can select several SWs. Then, TA will assign $token_{access}$ to each of them and store $token_{access}$ on the blockchain. Without the implicit access credential, the unselected SWs cannot acquire the *uiv* information. With the implicit access credential, a selected SW can only acquire the *uiv* information on the basis of his high trust value. In this way, the *uiv* sharing can be restricted within the scope of permission, and the leakage problem of dishonest SWs can be avoided in advance

In our UIV-TSP scheme, the metadata of *uiv* and the related SW access records are also stored on the blockchain, which can make the access logs of the *uiv* sharing as tamper-resistant. After the access, $token_{tracing}$ and $code_{preleak}$ can be stealthily sneaked into the acquired *uiv* information. Once the *uiv* information is one step away from the selected SW host, $code_{preleak}$ will destroy the *uiv* information and sends back a feedback containing the token, thus avoiding a widespread damage to the users of the IIoT manufacturing devices after the *uiv* leakage.

When the mitigation measures are developed, the *uiv* patch will be accurately distributed to the target IIoT manufacturing devices. Under the circumstances, the *uiv* information can be made public.

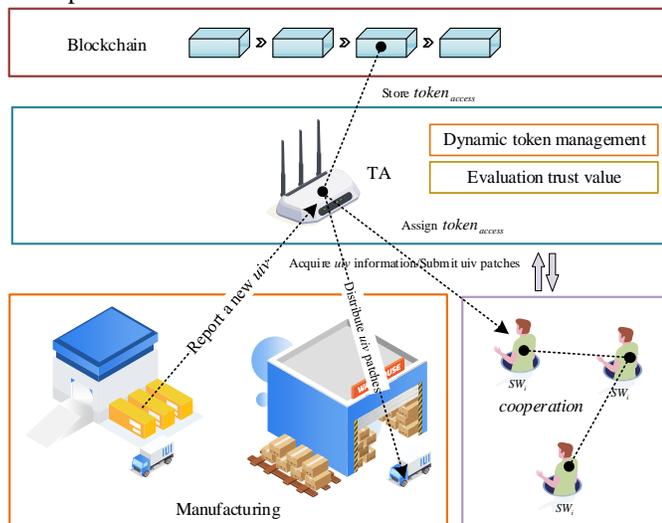

Fig. 12. UIV-TSP scheme in manufacturing.

## VI. CONCLUSION AND FUTURE WORK

In this article, we propose an undisclosed IoT vulnerabilities trusted sharing protection scheme with dynamic token. To facilitate the detection of leakage behaviors, we design a dynamic token as the implicit access credential and traceability clue. Assisted by blockchain, the continuous access logs of SWs can be secure storage. To prevent vulnerability leakage with one-step, we present a benign logic bomb called $code_{preleak}$ is embedded into the undisclosed IIoT vulnerability information. Trust mechanism is adopted to evaluate the trust value of SWs which can quickly distinguish SWs. Simulation results indicate that our proposed scheme is resilient to suppress dishonest SWs, and merely require limited traceability delay.

For future works, we will investigate on the selfish SWs and motivate them to develop the mitigation measures of undisclosed IIoT vulnerabilities under the protection of UIV-TSP.